\documentclass[%
 reprint,
 amsmath,amssymb,aps,
]{revtex4-2}
\usepackage{ulem}
\usepackage{appendix}
\usepackage{graphicx}
\usepackage{dcolumn}
\usepackage{bm}
\usepackage{color}

\begin{document}

\preprint{APS}

\title{Altermagnetic Weyl node-network metals protected by spin symmetry}

\author{Shuai Qu${^{1,2}}$}
\author{Xiao-Yao Hou${^{1,2}}$}
\author{Zheng-Xin Liu${^{1,2}}$}
\author{Peng-Jie Guo${^{1,2}}$}
\email{guopengjie@ruc.edu.cn}
\author{Zhong-Yi Lu${^{1,2,3}}$}
\email{zlu@ruc.edu.cn}
\affiliation{1. School of Physics and Beijing Key Laboratory of Opto-electronic Functional Materials $\&$ Micro-nano Devices. Renmin University of China, Beijing 100872, China}
\affiliation{2. Key Laboratory of Quantum State Construction and Manipulation (Ministry of Education), Renmin University of China, Beijing 100872, China}
\affiliation{3. Hefei National Laboratory, Hefei 230088, China}

\date{\today}

\begin{abstract}
Symmetry protected topology has been studied extensively in the past twenty years, but the topology protected by spin symmetry has just begun to be studied. 
In this work, based on spin symmetry analysis, we propose that a class of Weyl nodal line metals is protected by the spin symmetry.
Then, by the first-principles electronic structure calculations, we predict that both altermagnetic $\rm Nb_2FeB_2$ and $\rm Ta_2FeB_2$ are node-network metals protected by the spin symmetry. Moreover, both altermagnetic $\rm Nb_2FeB_2$ and $\rm Ta_2FeB_2$ have nodal rings protected by the mirror symmetry and Dirac points protected by nonsymmorphic spin symmetry. Furthermore, both altermagnetic $\rm Nb_2FeB_2$ and $\rm Ta_2FeB_2$  transform node-network metal phase into Weyl metal phase when considering spin-orbit coupling. Therefore, our work not only enriches the topological phases protected by spin symmetry, but also provides an excellent material platform to investigate the exotic physical properties of multiple altermagnetic topological metal phases in experiment.
\end{abstract}

\maketitle


{\it Introduction.} Symmetry protected topology (SPT) has been studied extensively in condensed matter physics due to rich topological phases and novel physical properties\cite{RMP-Hasan, RMP-Qi, Weng_2016, addNM1,addNM2, nature-TQC, NC-SI-2017, NC-song, NP-Tang, Nature-Tang, Nature-wang, Nature-zhang, Nature-Xu, NC-Xu, Peng-PRB, addAFM,Yao-MSG-1,Yao-MSG-2}. 
Symmetry plays a pivotal role in the classification of topology and the study of topological properties of materials. 
Nonmagnetic materials with negligible and strong spin-orbital coupling (SOC) are described by type-II magnetic space groups (MSGs) and their corresponding double groups, respectively. For nonmagnetic materials, the topological classification has been completed by the theories of the symmetry-based indicator and topological chemistry based on band representations, leading to a dictionary of nonmagnetic topological materials\cite{nature-TQC, NC-SI-2017, NC-song, NP-Tang, Nature-Tang, Nature-wang, Nature-zhang,addNM1,addNM2}. 
On the other hand, the magnetic materials with SOC are described by type-I, II, and IV double MSGs. In consequence, the topological classification has also been completed by the theory of magnetic topological chemistry\cite{Nature-Xu, NC-Xu, Peng-PRB, addAFM,Yao-MSG-1,Yao-MSG-2}. Moreover, many magnetic topological materials have been extensively studied by theory and experiment, such as $\rm MnBi_2Te_4$\cite{Nature-MnBiTe}.

The magnetic materials with negligible SOC can be described by spin space groups. 
The spin space groups represent the symmetry transformation of decoupled spin and lattice space. 
Due to decoupled spin and lattice space, the symmetry transformation of spin and lattice space can be distinct rotation operations. 
Therefore, conventional MSGs are only subgroups of spin space groups (SSGs). Since the symmetry landscape of SSGs is more plentiful than that of conventional MSGs, there are more novel topological phases protected by spin symmetry in condensed matter physics. Recently, some topological metal and insulator phases protected by spin symmetry have been proposed in theory\cite{Guo-PRL, Liu-PRX, Liu-In}. Moreover, full classification on spin space group has also been completed\cite{Spin-PRX1, Spin-PRX2, Spin-PRX3}. These works inspire further investigation into the exotic topological phases of spin symmetry protection.

Recently, based on spin group theory, altermagnetism different from ferromagnetism and antiferromagnetism (AFM) has been proposed\cite{altermagnetism-PRX-1, altermagnetism-PRX-2}. In altermagnetism, opposite spin sublattices cannot be connected by space-inversion (I) and fractional translation ($\tau$) operations, but only by rotation or mirror (R) operations. Correspondingly, altermagnetic materials break the spin symmetry $\{{\rm C_2^\bot || I}\}$ and $\{{\rm C_2^\bot || \tau}\}$, but have the spin symmetry $\{{\rm C_2^\bot || R}\}$ (the $\rm C_2^\perp$ represents 180 degrees rotation perpendicular to the spin direction). The breaking of the spin symmetries $\{{\rm C_2^\bot || I}\}$ and $\{{\rm C_2^\bot || \tau}\}$ results in the spin splitting, while the spin symmetry $\{{\rm C_2^\bot || R}\}$ guarantees the total magnetic moment to be zero. Therefore, altermagnetism has the duality of real-space antiferromagnetism and reciprocal-space anisotropic spin splitting. On the other hand, under spin-orbit coupling, the symmetry of altermagnetic materials will change from spin group symmetry to magnetic group symmetry. In particular, the spin symmetries $\{{\rm C_2^\bot || I}\}$ and $\{{\rm C_2^\bot || \tau}\}$ will change to IT and T$\tau$ symmetries (equivalent time-reversal symmetries), respectively, so that the altermagnetic materials break the equivalent time-reversal symmetries. Due to the duality of real-space antiferromagnetism and reciprocal-space anisotropic spin splitting and the breaking of equivalent time-reversal symmetry under spin-orbit coupling, altermagnetic materials have many novel physical properties\cite{altermagnetism-PRX-3, SST-PRL2021, SST-PRL2022, SST-NE2022, SST-PRL2022-2, GMR-PRX2022, TMR-Shao2021, piezomagnetism-NC, AHE-Sinova2022, AHE-RuO2-NE2022, AHE-hou2023, AHE-MnTe-PRL2023, QAH-npj2023, SC-AM, CTHE-Yao2024, LiFe2F6-guo2023, NiF3-qu2024, MCM-liu2023, HighoT-liu2024, AI-gao2023, YJZ-2024, AM-nature2024, liu-arxiv2024, AM-CrSb2024, AM-MnTe2024, Tan2024}. Therefore, altermagnetic materials may be an excellent platform for exploring novel topological phases.

In this work, based on spin symmetry analysis, we propose a new Weyl nodal line phase protected by the spin symmetry $\{\rm{TC_2^\perp||IT}\}$. 
Then, by the first-principles electronic structure calculations, we predict that both altermagnetic $\rm Nb_2FeB_2$ and $\rm Ta_2FeB_2$ are Weyl node-network metals protected by the spin symmetry $\{\rm {TC_2^\perp||IT}\}$. Moreover, both $\rm Nb_2FeB_2$ and $\rm Ta_2FeB_2$ also have Weyl nodal lines protected by the mirror symmetry and 16 Dirac points protected by nonsymmorphic spin symmetry. When considering SOC, both altermagnetic $\rm Nb_2FeB_2$ and $\rm Ta_2FeB_2$ will transform the node-network metal phase into the Weyl metal phase.

{\it Method} The first-principles electronic structure calculations were performed in the framework of density functional theory (DFT) using the Vienna Abinitio Simulation Package (VASP)\cite{DFT1,DFT2,Vasp1,Vasp2,Vasp3}. 
The generalized gradient approximation (GGA) of Perdew–Burke–Ernzerhof (PBE) type was adopted for the exchange correlation functional\cite{GGAPBE}. 
The projector augmented wave (PAW) method was adopted to describe the interactions between valence electrons and nuclei\cite{Paw1,Paw2}. 
The kinetic energy cutoff of the plane-wave basis was set to be $\rm 600 eV$. 
The total energy convergence and the atomic force tolerance were set to be $\rm 10^{-6} eV$ and $\rm 0.01 eV/\AA$, respectively. 
For describing the Fermi-Dirac distribution function, a Gaussian smearing of $\rm 0.05 eV$ was used. An $8\times8\times15$ k-point mesh was used for the Brillouin Zone (BZ) sampling. 
To account for the correlation effects of Fe $\bm{3d}$ orbitals, we performed $\rm GGA+U$ calculations by using the simplified rotationally invariant version of Dudarev et al\cite{LDAU1,LDAU2}. 
By the linear response method\cite{LinearU}, the onsite effective $\rm U_{eff}$ values of $\rm {Fe}$ $\bm{d}$-electron of $\rm Nb_2FeB_2$ and $\rm Ta_2FeB_2$ were estimated to be $\rm 4.82 eV$ and $\rm 4.76 eV$, respectively, which were used to study electronic structures of $\rm Nb_2FeB_2$ and $\rm Ta_2FeB_2$. The topological properties of $\rm Nb_2FeB_2$ and $\rm Ta_2FeB_2$ were calculated by \rm\texttt{wannier90}\cite{Wannier90,Wannier90-1,Wannier90-2} and $\rm\texttt{Wanniertools}$\cite{WU2017} software packages.

{\it Symmetry analysis}
For spin point groups, since spin space and lattice space are decoupled, the symmetric operations on spin space and lattice space can be different or the same, such as  $\{\rm {C_{2x}||C_{2y}}\}$ and  $\{\rm {C_{2x}||C_{2x}}\}$. For magnetic point groups, the spin space and the lattice space are coupled, so that the symmetric operations on the spin space and the lattice space must be the same, such as $\{\rm {C_{2x}||C_{2x}}\}$. Therefore, the magnetic point group is a subgroup of the corresponding spin point group. On the other hand, any collinear magnetism (including ferromagnetism and antiferromagnetism) has spin symmetry $\{{\rm TC_2^\bot||T}\}$, and if the corresponding material also has spin symmetry $\{{\rm E||I}\}$, then the material must have spin symmetry $\{\rm {TC_2^\perp||IT}\}$. Although the time-reversal symmetry $\rm T$ does not have an effect on the lattice space, the effect on the reciprocal space can change $\bm k$ to $-\bm{k}$. Combined with space-inversion symmetry $\rm I$ in reciprocal space, every $\bm k$ point has the spin symmetry $\{\rm {TC_2^\perp||IT}\}$.
In two-dimensional systems with the spin symmetry $\{\rm {TC_2^\perp||IT}\}$, the Berry phase for loops having fully gapped spectrum is quantized to 0 or $\pi$ depending on the number of Weyl points to be even or odd\cite{QAH-npj2023}. 
That is to say, the spin symmetry $\{\rm {TC_2^\perp||IT}\}$ can protect Weyl points at general $\rm \pmb{k}$ points in the two-dimensional BZ\cite{QAH-npj2023}. If generalizing two-dimensional systems to three-dimensional systems, the spin symmetry $\{\rm {TC_2^\perp||IT}\}$ can naturally protect three-dimensional Weyl node-line metals. Moreover, the Weyl nodal line can be located in any region of BZ, which is similar to the Dirac nodal line protected by IT symmetry in nonmagnetic materials.
Then, a natural question is what materials may be Weyl node-line metals protected by the spin symmetry $\{\rm {TC_2^\perp||IT}\}$. First, the electronic bands of these materials have spin-splitting in the nonrelativistic case. Second, these materials have the spin symmetry $\{\rm {TC_2^\perp||IT}\}$ and the corresponding electronic structures have band crossing. 

{\it Results} Very recently, $\rm Nb_2FeB_2$ and $\rm Ta_2FeB_2$ have been predicted to be $\bm g$-wave altermagnetic materials with large intrinsic anomalous Hall effects (AHE)\cite{AHE-hou2023, AI-gao2023}. 
The crystal structures of $\rm Nb_2FeB_2$ and $\rm Ta_2FeB_2$ are illustrated in Fig. 1(a). From the Fig. 1(a), the crystal primitive cell contains two formula units and is composed of $\rm Fe-B$ and $\rm Nb(Ta)$ atomic layers. 
The magnetic primitive cell is the same as the crystal primitive cell (Fig. 1(a)) and the corresponding bulk BZ and projected two-dimensional surface BZ are illustrated in Fig. 1(b). In the nonrelativistic case, the symmetry of altermagnetic $\rm Nb_2FeB_2$ and $\rm Ta_2FeB_2$ are described by spin group and the corresponding generators are $\{\rm {TC_2^\perp||T}\}$, $\{\rm{E||I}\}$, $\{\rm{C_2^\perp||C_{2x}(1/2, 1/2)}\}$, $\{\rm{E||C_{2z}}\}$ and $\{\rm{E||C_{4z}}\}$. 
Therefore, both $\rm Nb_2FeB_2$ and $\rm Ta_2FeB_2$ have the spin symmetry $\{\rm {TC_2^\perp||IT}\}$. 

\begin{figure}[htbp]
\includegraphics[width=0.38\textwidth]{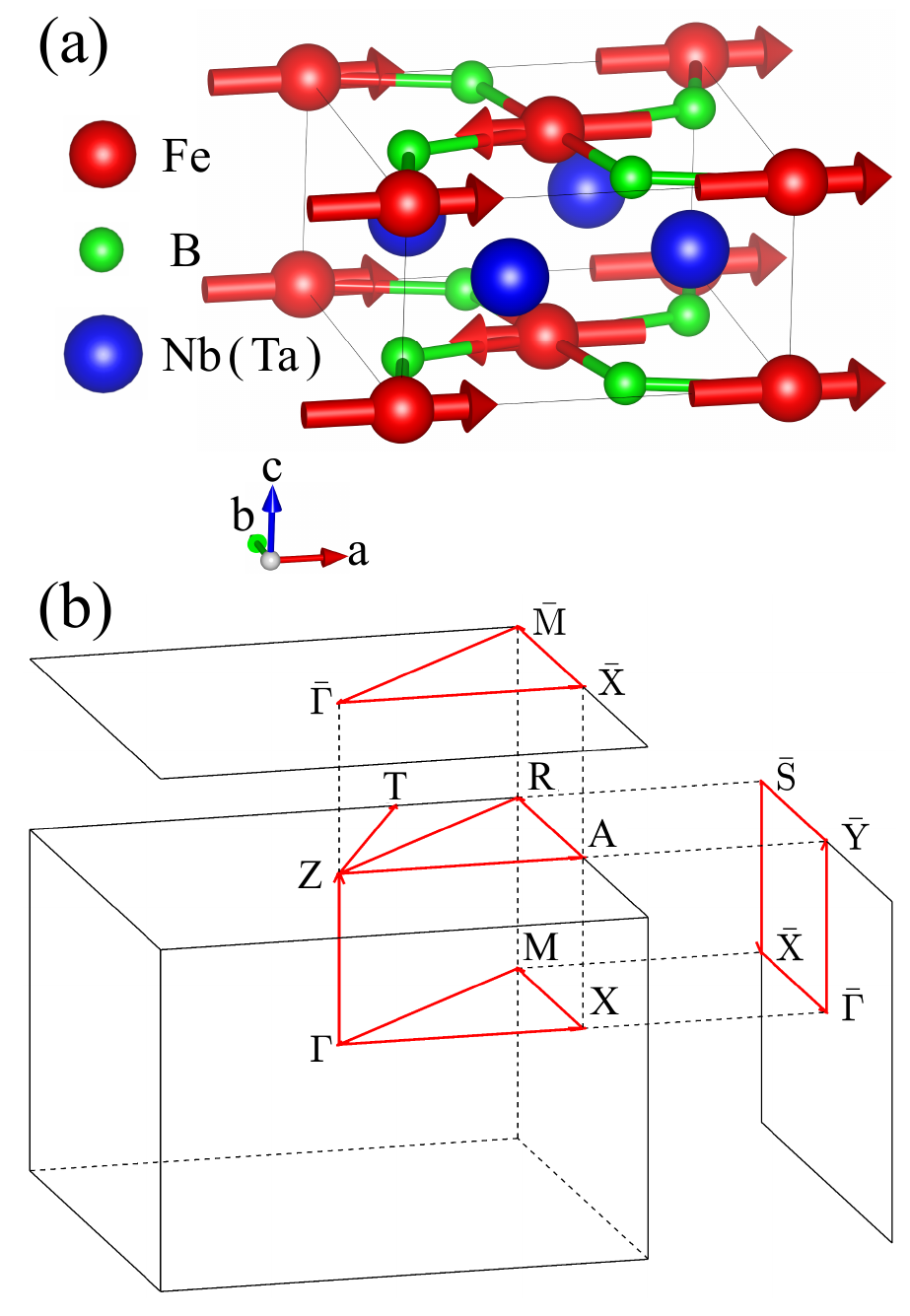}
\caption{
(a) Crystal structures of the $\rm Nb_2FeB_2$ and $\rm Ta_2FeB_2$. 
The red arrows indicate magnetic momentum direction. 
(b) The corresponding bulk BZ and projected two-dimensional surface BZ. 
The red lines represent the high-symmetry paths.}
\label{f1}
\end{figure}

Now we study the electronic structures of altermagnetic $\rm Nb_2FeB_2$ and $\rm Ta_2FeB_2$. Since altermagnetic $\rm Nb_2FeB_2$ and $\rm Ta_2FeB_2$ have the similar electronic structures, we only show the results of $\rm Nb_2FeB_2$ in the following. 
The calculated electronic band structures of altermagnetic $\rm Nb_2FeB_2$ without SOC are shown in Fig. 2(a). From Fig. 2(a), the $\rm Nb_2FeB_2$ is an altermagnetic metal. Due to the absence of the spin symmetries $\{{\rm C_2^\bot || I}\}$ and $\{{\rm C_2^\bot || \tau}\}$, spin-up and spin-down bands should be split in BZ of the $\rm Nb_2FeB_2$. However, these bands of $\rm Nb_2FeB_2$ along the high-symmetry directions are always degeneracy. By symmetry analysis, we find that the spin degeneracy along the high-symmetry direction is protected by these spin symmetries $\{\rm{C_2^\perp||C_{2x}(1/2, 1/2)}\}$, $\{\rm{C_2^\perp||C_{2y}(1/2, 1/2)}\}$, $\{\rm{C_2^\perp||C_{xy}(1/2, 1/2)}\}$ and $\{\rm{C_2^\perp||C_{x-y}(1/2, 1/2)}\}$. To demonstrate the anisotropic spin splitting caused by altermagnetism, we also calculate the band structures of $\rm Nb_2FeB_2$ along the non-high-symmetry Z-T direction. Indeed, these bands along the non-high-symmetry Z-T direction are spin-splitting reflecting the characteristics of $\bm g$-wave splitting (Fig. 2(a)).


\begin{figure}[htbp]
\includegraphics[width=0.49\textwidth]{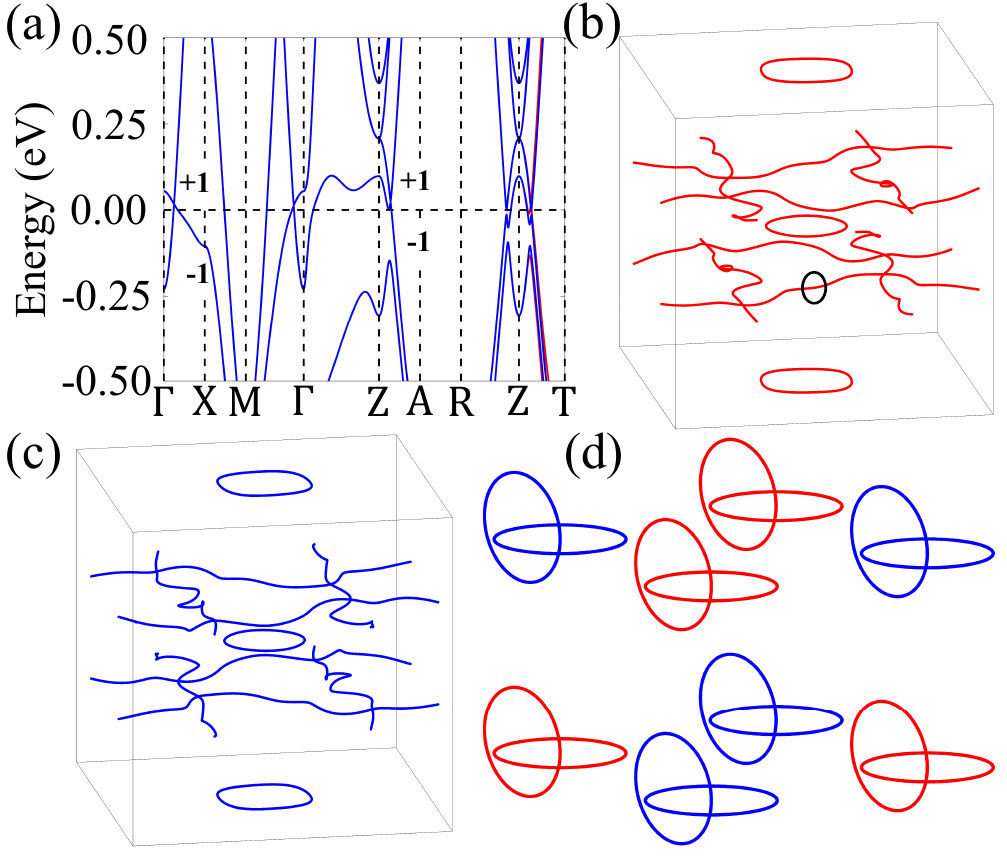}
\caption{
(a) The spin-resolved electronic band structures of $\rm Nb_2FeB_2$ without SOC along the high-symmetry and non-high-symmetry directions. 
(b) and (c) are nodal line structures of spin-up and spin-down bands, respectively. 
(d) Hopf-link nodal lines protected by the spin symmetry $\{\rm{TC_2^\perp||IT}\}$. The red and blue represent the spin-up and spin-down bands, respectively. 
The $1$ and $-1$ are the eigenvalues of the mirror $\{\rm{E||M_z}\}$ of the two crossing bands around the Fermi level. The black ring represents a closed path around the nodal line.}
\label{f2}
\end{figure}

Interestingly, altermagnetic $\rm Nb_2FeB_2$ has multiple crossing points around the Fermi level, which implies altermagnetism with nontrivial topological properties. These crossing points on the $\rm \Gamma-X$, $\rm \Gamma-M$, $\rm Z-A$, $\rm Z-R$ and $\rm Z-T$ axes indicate that altermagnetic $\rm Nb_2FeB_2$ may have two Weyl nodal rings protected by the mirror symmetry $\{\rm {E||M_z}\}$ on $\rm k_z = 0$ and $\pi$ planes around the Fermi level for spin-up or spin-down bands (Fig. 2(a)). 
To confirm it, we calculate nodes of two crossing spin-up (spin-down) bands around the Fermi level in the whole BZ, which are shown in Fig. 2(b) and (c). Obviously, there are two Weyl nodal rings on the $\rm k_z = 0$ and $\pi$ planes for spin-up (spin-down) bands. Since the two crossing bands with opposite eigenvalues of mirror $\{\rm {E||M_z}\}$ (Fig. 2(a)), the two Weyl nodal rings on the $\rm k_z = 0$ and $\pi$ planes are protected by the mirror symmetry $\{\rm {E||M_z}\}$ . On the other hand, due to the spin degeneracy along the high-symmetry directions, all crossing points on the $\rm \Gamma-X$, $\rm \Gamma-M$, $\rm Z-A$ and $\rm Z-R$ are fourfold degenerate Dirac points. Considering the spin symmetry $\{\rm {E||C_{4z}}\}$, altermagnetic $\rm Nb_2FeB_2$ has 16 Dirac points.    

More significantly, in addition to the two Weyl nodal rings on the $\rm k_z = 0$ and $\pi$ planes, there are two Weyl nodal networks between $\rm k_z=0$ and $\pi$ for spin-up (Fig. 2(b)) or spin-down (Fig. 2(c)) bands. 
Since located at general area in the BZ, the two Weyl nodal networks are only protected by the spin symmetry $\{\rm{TC_2^\perp||IT}\}$ . 
As far as we know, the class of new Weyl nodal line metal is proposed for the first time.
Moreover, the Weyl nodal network consists of four nodal lines and the four nodal lines don't intersect with each other.  To characterize the topology of the node-network, we calculate the integral of the Berry curvature along a closed path around the nodal line (Fig. 2(b)). The Berry phase along a closed path around the nodal line is $\pi$, which indicates the nontrivial topology of node-network metal $\rm Nb_2FeB_2$. Due to periodic boundary condition, a Weyl nodal network can form two Hopf-link Weyl nodal lines (Fig. 2(d))\cite{Hopf-link}. Therefore, there are eight Hopf-link Weyl nodal lines for both spin-up and spin-down bands in Fig. 2(d). In addition, the calculations using the SCAN and HSE functionals also indicate that $\rm Nb_2FeB_2$ possesses the same topological properties.

\begin{figure}[htbp]
\includegraphics[width=0.49\textwidth]{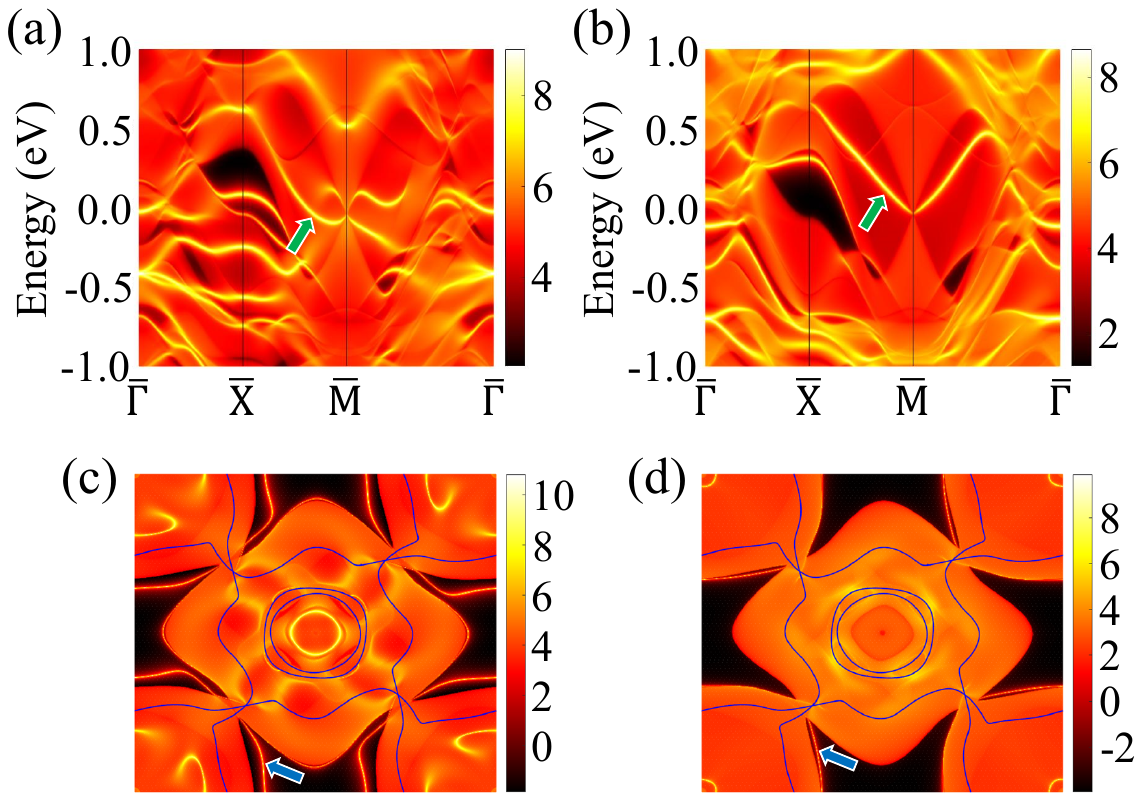}
\caption{
Spectral function of unconventional collinear AFM $\rm Nb_2FeB_2$ along the high-symmetry directions in the projected 2D BZ of the (001) surface with left (a) and right (b) semi-infinite termination boundaries for the spin-up bands. 
The spectral function at the Fermi level with left (c) and right (d) semi-infinite termination boundaries. 
The blue lines represent nodal lines of $\rm Nb_2FeB_2$ in the absence of SOC.}
\label{f3}
\end{figure}

As well-known to all, nontrivial node-line metals have topological protected drumhead surface states. Since these Weyl nodal networks and rings can be projected to the $(001)$ surface, altermagnetic $\rm Nb_2FeB_2$ has nontrivial drumhead surface states on the $(001)$ surface, which are confirmed by the calculated spectral function with semi-infinite termination boundary. Since spin-up and spin-down electronic structures have similar topological surface states, we only show these results for spin-up electronic structures. From the Fig. 3(a) and (b), there are nontrivial drumhead surface states derived from the nodal rings protected by mirror symmetry, which are marked by green arrows. Interestingly, the nodal networks protected by the spin symmetry $\{\rm {TC_2^\perp||IT}\}$ also have nontrivial surface states which are marked by blue arrows in Fig. 3(c) and (d).

\begin{figure}[htbp]
\includegraphics[width=0.5\textwidth]{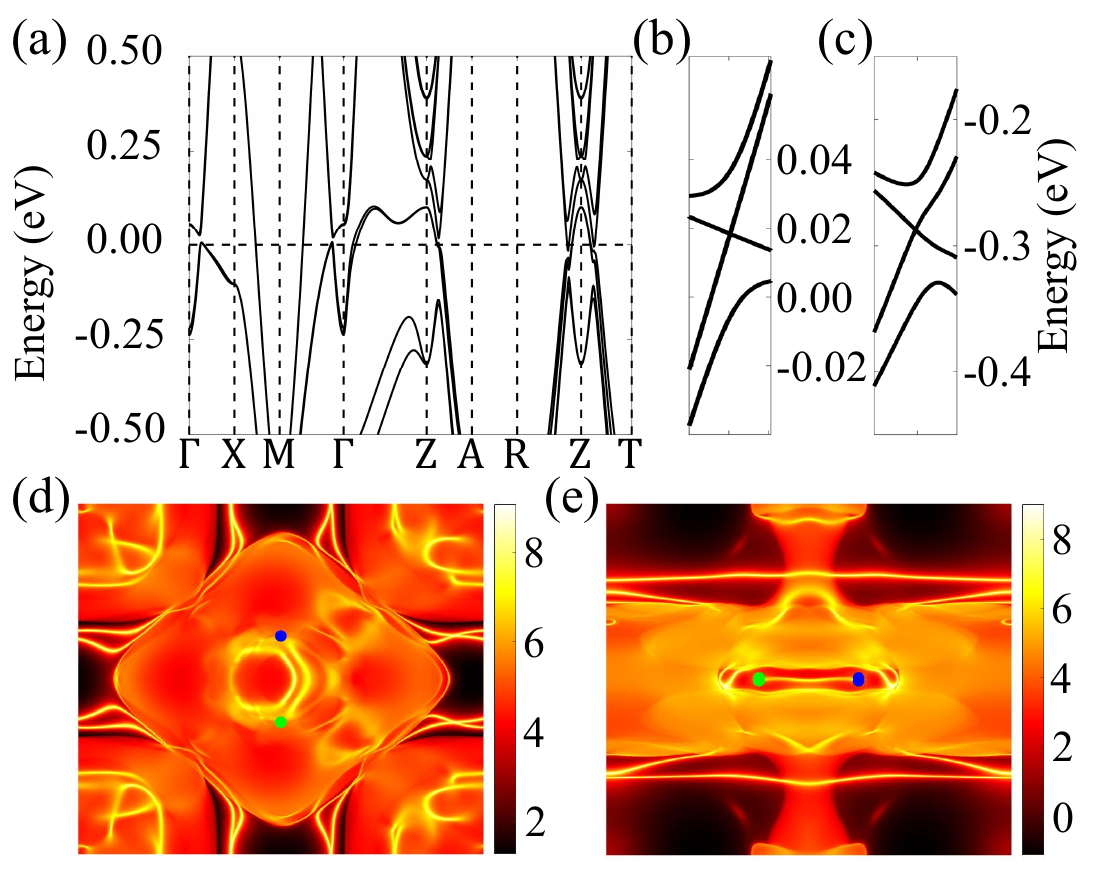}
\caption{
(a) The electronic band structures of unconventional collinear AFM $\rm Nb_2FeB_2$ with SOC along the high-symmetry and non-high-symmetry directions. 
(b) and (c) represent Weyl points above and under the Fermi level, respectively. 
The spectral function at the energy of Weyl points with semi-infinite termination at the (d) (001) and (e) (100) boundaries. 
The green and blue dots represent Weyl points with topological charge 1 and -1, respectively.}
\label{f4}
\end{figure}

With SOC, due to coupled spin and lattice space, the symmetry transformation of spin and lattice space must be the same symmetry operation. Correspondingly, the symmetry of altermagnetic $\rm Nb_2FeB_2$ changes from spin point group to magnetic point group. Moreover, the magnetic point group symmetry depends on the direction of the easy magnetization axis. Since the easy magnetization axis of altermagnetic $\rm Nb_2FeB_2$ is along the $\bm x$ direction\cite{AHE-hou2023}, altermagnetic $\rm Nb_2FeB_2$  has $\rm C_{2y}(1/2, 1/2)$, $\rm I$, $\rm M_y(1/2, 1/2)$, $\rm C_{2x}(1/2, 1/2)T$, $\rm M_x(1/2, 1/2)T$, $\rm C_{2z}T$ and $\rm M_zT$ symmetries. 
Then, we calculate the electronic band structures of altermagnetic $\rm Nb_2FeB_2$ with SOC, which are illustrated in Fig. 4 (a). Since the spin symmetries $\{\rm {E||Mz}\}$ and $\{\rm {TC_2^\perp||IT}\}$ are broken, these Weyl nodal rings and networks open gap, but the bandgap is very small due to the weak SOC (Fig. 4 (a)), such as the bandgap being $\rm 18meV$ and $\rm 10meV$ on the $\Gamma$-X and $\Gamma$-M directions, respectively. 
This is advantageous to the study of the novel properties of nodal networks protected by the spin symmetry $\{\rm {TC_2^\perp||IT}\}$ in experiment. 

\begin{table}
\caption{The position and chirality of 12 independent Weyl points in the BZ of altermagnetic $\rm Nb_2FeB_2$.}
\label{TaBS1}
\begin{ruledtabular}
\begin{tabular}{ccrc}
&Position & Chirality& \\ \hline
&(0.492, 	0.484, 	0.041) 	&-1.000 &\\
&(0.259, 	0.445, 	0.148) 	&-1.000 &\\
&(0.391, 	0.270, 	0.137) 	&1.000 &\\
&(0.467, 	0.252, 	0.156) 	&-1.000 &\\
&(0.000, 	0.123, 	0.004) 	&-1.000 &\\
&(0.181, 	0.222, 	0.117) 	&1.000 &\\
&(0.213, 	0.123, 	0.159) 	&-1.000 &\\
&(0.118, 	0.213, 	0.157) 	&1.000 &\\
&(0.215, 	0.109, 	0.150) 	&1.000 &\\
&(0.000, 	0.280, 	0.127) 	&-1.000 &\\
&(0.271, 	0.021, 	0.124) 	&-1.000 &\\
&(0.000, 	0.259, 	0.121) 	&-1.000 &
\end{tabular}
\end{ruledtabular}
\end{table}

In general, Weyl node-line metals transform into Weyl metals when considering SOC. Our calculations also show that altermagnetic $\rm Nb_2FeB_2$ indeed transforms into a Weyl metal phase and has 84 Weyl points. According to magnetic symmetry analysis, there are only 12 independent Weyl points, which are shown in Table I. Then, we calculate the two Weyl points closest to the Fermi level for altermagnetic $\rm Nb_2FeB_2$ which are shown in Fig. 4 (b) and (c). From Fig. 4 (b) and (c), altermagnetic $\rm Nb_2FeB_2$ is type-I Weyl metal. In order to display the characteristics of Weyl metal more intuitively, we also calculate the chirality of all Weyl points (Table I) and topological protected Fermi arc in Fig. 4 (d) and (e). From Fig. 4 (d) and (e), topological protected Fermi arc connects the two Weyl points with opposite chirality. Thus, the $\rm Nb_2FeB_2$ is an altermagnetic Weyl metal under SOC effect. Finally, our calculations also show that $\rm Ta_2FeB_2$ has the same topological properties as $\rm Nb_2FeB_2$.

In summary, based on the symmetry analysis, we propose a new type of Weyl node-line metal protected by the spin symmetry $\{\rm {TC_2^\perp||IT}\}$. Then, by the first-principles electronic structure calculations, we predict that altermagnetic $\rm Nb_2FeB_2$ and $\rm Ta_2FeB_2$ are node-network metal protected by the spin symmetry $\{\rm {TC_2^\perp||IT}\}$. 
Moreover, both altermagnetic $\rm Nb_2FeB_2$ and $\rm Ta_2FeB_2$ will transform the node-network metal phase into the Weyl metal phase when considering SOC. 
Our study not only enriches the topology class protected by the spin symmetry, but also provides an excellent platform to explore the exotic physical properties of node-network metals protected by the spin symmetry.

\begin{acknowledgments}
This work was financially supported by the National Natural Science Foundation of China (No.12204533 and No.12434009), the Fundamental Research Funds for the Central Universities, and the Research Funds of Renmin University of China (Grant No. 24XNKJ15) and the Innovation Program for Quantum Science and Technology (Grants No. 2021ZD0302402). Computational resources have been provided by the Physical Laboratory of High Performance Computing at Renmin University of China.
\end{acknowledgments}

\appendix
\begin{appendices}
\section{Topological properties of altermagnetic material $\rm Ta_2FeB_2$}

As demonstrated in Fig. 5(a), $\rm Ta_2FeB_2$ is also an altermagnetic metal. Due to the absence of the spin symmetries $\rm \{C_2^\perp||I\}$ and $\rm \{C_2^\perp||\tau\}$, the spin-up and spin-down bands are split along the non-high-symmetry $\rm Z-T$ direction without SOC, which reflects the $\bm g$-wave characteristics in the BZ of $\rm Ta_2FeB_2$.
From Fig. 5(b) and (c), there are nodal rings and networks in BZ of the $\rm Ta_2FeB_2$ without SOC. The Weyl nodal rings on the $\rm k_z=0$ and $\pi$ planes are protected by the mirror symmetry $\rm \{E||M_z\}$.
In consideration of the rotation symmetry $\rm \{E||C_{4z}\}$ and spin degeneracy along the high-symmetry directions, altermagnetic metal $\rm Ta_2FeB_2$ also has 16 Dirac points around the Fermi level, such as the crossing point in the $\rm \Gamma-X$ direction.

\begin{figure}[htbp]
\includegraphics[width=0.5\textwidth]{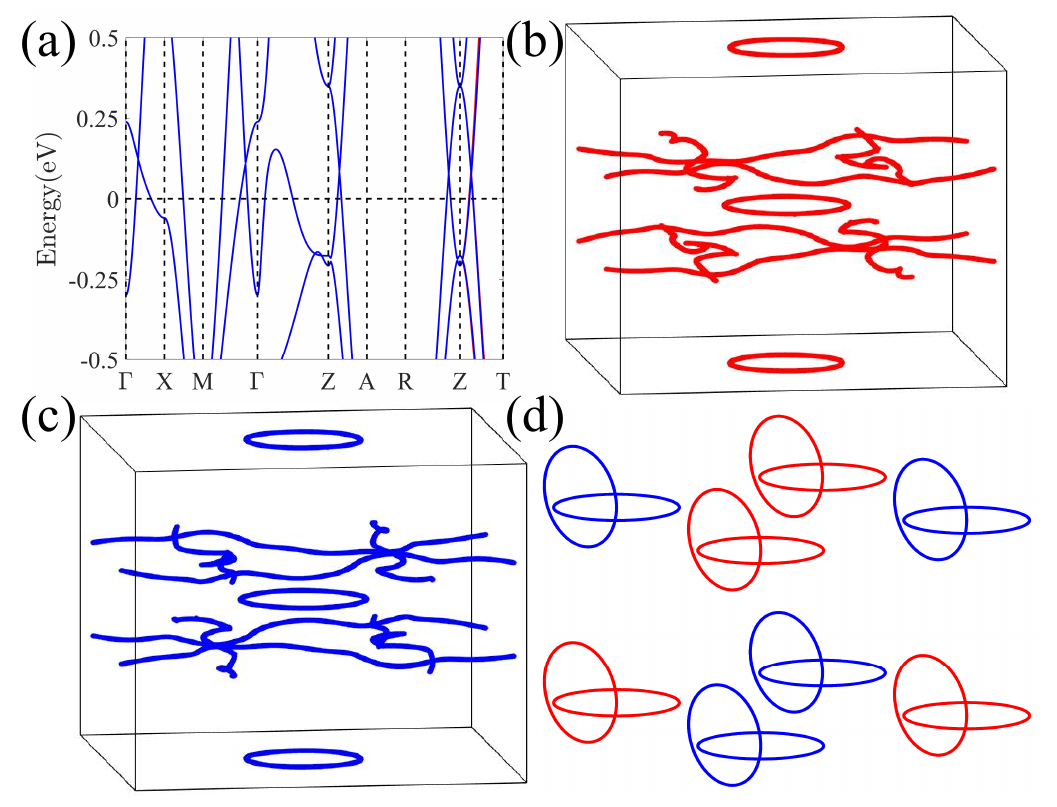}
\caption{
(a) The spin-resolved electronic band structures of altermagnetic $\rm Ta_2FeB_2$ without SOC along the high-symmetry and non-high-symmetry directions. 
(b) and (c) are nodal line structures of spin-up (red lines) and spin-down (blus lines) bands, respectively. 
(d) The schematic diagram of eight Hopf-link nodal lines protected by the spin symmetry $\{\rm{TC_2^\perp||IT}\}$ for both the spin-up and spin-down bands. The red and blue lines represent the spin-up and spin-down band crossings, respectively. }
\end{figure}

The two Weyl nodal line networks within the general BZ for spin-up band crossings in Fig. 5(b), as well as spin-down band crossings in Fig. 5(c), are only protected by the spin symmetry $\rm \{TC_2^\perp||IT\}$.
The Weyl nodal line network consists of four nodal lines that do not intersect with each other. Nevertheless, a Weyl nodal line network could form two Hopf-link Weyl nodal lines due to periodic boundary condition. Thus, there are eight Hopf-link Weyl nodal lines for both spin-up and spin-down bands in Fig. 5(d).

\begin{figure}[htbp]
\includegraphics[width=0.5\textwidth]{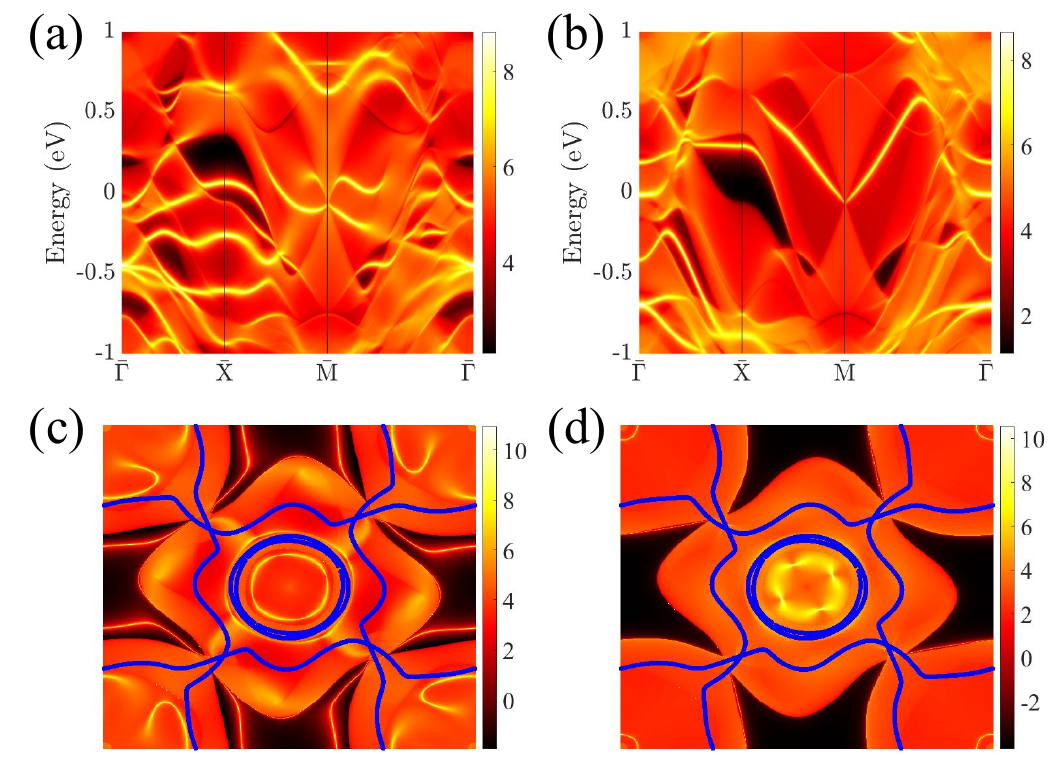}
\caption{
Spectral function of unconventional collinear AFM $\rm Ta_2FeB_2$ along the high-symmetry directions in the projected 2D BZ of the (001) surface with left (a) and right (b) semi-infinite termination boundaries for the spin-up bands. 
The spectral function at the Fermi level with left (c) and right (d) semi-infinite termination boundaries. 
The blue lines represent Weyl nodal lines in the BZ of altermagnetic $\rm Ta_2FeB_2$ without SOC.}
\end{figure}

These nodal rings and networks can lead to nontrivial surface states on boundary for $\rm Ta_2FeB_2$. Since the Weyl nodal rings and networks could be projected to the (001) surface, altermagnetic metal $\rm Ta_2FeB_2$ has nontrivial drumhead surface states on the (001) surface, which are confirmed by the calculated spectral function with semi-infinite termination boundary. As illustrated in Fig. 6(a) and (b), there are nontrivial drumhead surface states derived from the nodal rings protected by mirror symmetry $\rm \{E||M_z\}$ for spin-up band structures.
Interestingly, the nodal networks protected by the spin symmetry $\rm \{TC_2^\perp||IT\}$ also have nontrivial surface states in Fig. 6(c) and (d).

\begin{figure}[htbp]
\includegraphics[width=0.5\textwidth]{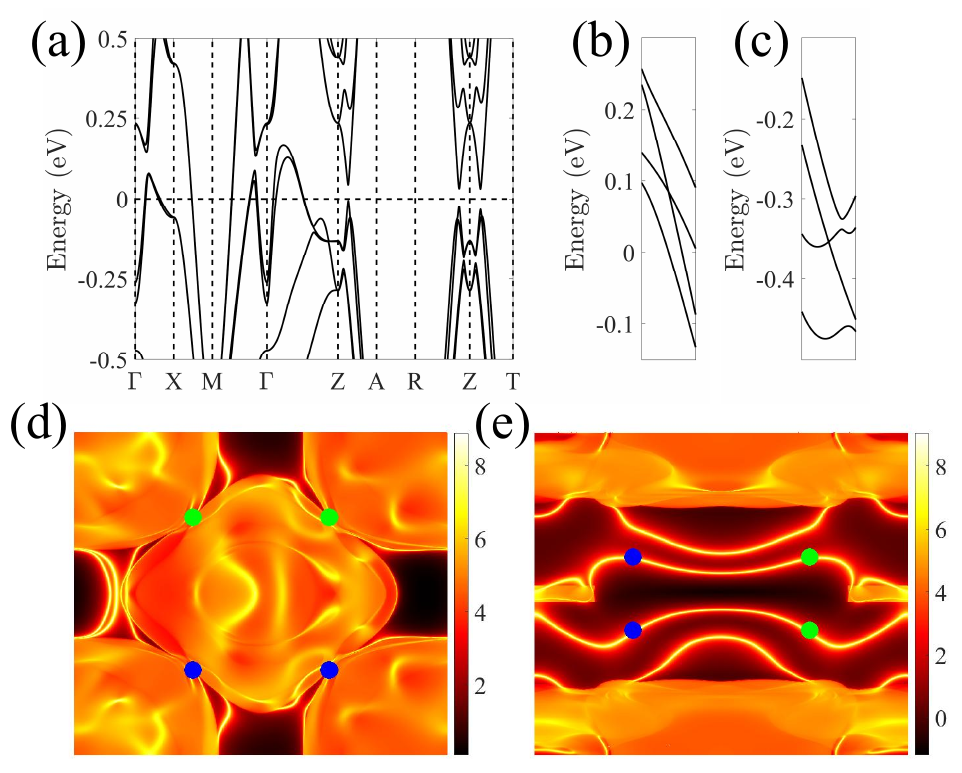}
\caption{
(a) The electronic band structures of unconventional collinear AFM $\rm Ta_2FeB_2$ with SOC along the high-symmetry and non-high-symmetry directions. 
(b) and (c) represent Weyl points above and under the Fermi level, respectively. 
The spectral function at the energy of the Weyl points above the Fermi level, with semi-infinite termination at the (d) (001) and (e) (100) boundaries. 
The green and blue dots represent Weyl points with topological charge 1 and -1, respectively.}
\end{figure}

When considering SOC, the spin symmetries $\rm \{E||M_z\}$ and $\rm \{TC_2^\perp||IT\}$ are broken, and these Weyl nodal rings and networks open gap. As illustrated in Fig. 7(a), the SOC effect in $\rm Ta_2FeB_2$  is stronger than that in $\rm Nb_2FeB_2$, such that the bandgaps are equal to 90$\rm meV$ and 64$\rm meV$ in the $\rm \Gamma-X$ and $\rm \Gamma-M$ directions, respectively. Moreover, the altermagnetic $\rm Ta_2FeB_2$ transforms into a Weyl metal phase with 72 Weyl points in the BZ which are shown in Table II. The two Weyl points closest to the Fermi level of altermagnetic metal $\rm Ta_2FeB_2$ are shown in Fig. 7(b) and (c). From Fig. 7(b) and (c), altermagnetic $\rm Ta_2FeB_2$ contains both Type-I and Type-II Weyl fermions.

\begin{table}
\caption{The position and chirality of 11 independent Weyl points in the BZ of altermagnetic $\rm Ta_2FeB_2$.}
\label{TaBS2}
\begin{ruledtabular}
\begin{tabular}{ccrc}
&Position & Chirality& \\ \hline
&(0.000, 	0.282, 	0.121) 	&-1.000&\\ 
&(0.000, 	0.239, 	0.115) 	&-1.000&\\ 
&(0.000, 	0.154, 	0.017) 	&-1.000&\\ 
&(0.096, 	0.206, 	0.141) 	& 1.000&\\ 
&(0.183, 	0.237, 	0.113) 	& 1.000&\\ 
&(0.191, 	0.171, 	0.112) 	& 1.000&\\ 
&(0.193, 	0.172, 	0.113) 	&-1.000&\\ 
&(0.251, 	0.067, 	0.121) 	&-1.000&\\ 
&(0.258, 	0.447, 	0.133) 	&-1.000&\\ 
&(0.485, 	0.478, 	0.048) 	&-1.000&\\ 
&(0.496, 	0.485, 	0.000) 	& 1.000& 
\end{tabular}
\end{ruledtabular}
\end{table}


\end{appendices}
\newpage


\bibliography{main.bib}

\end{document}